\documentclass[aps,pre,twocolumn,showpacs,superscriptaddress,showkeys,showpacs,groupedaddress]{revtex4}
\usepackage{graphics}
\usepackage{amsmath}
\usepackage{amssymb}
\usepackage{amsfonts}
\usepackage{graphicx}
\usepackage{color}
\usepackage{soul}
\usepackage[caption=false]{subfig}
\usepackage[toc,page]{appendix}

\graphicspath{{paper_plots/}}

\begin{document}
%
% \title{Chimera patterns in three-dimensional networks of coupled Leaky Integrate-and-Fire oscillators}
\title{Three-dimensional chimera patterns in networks of spiking neuron oscillators}

\author{T. Kasimatis}
\affiliation{Institute of Nanoscience and Nanotechnology, National Center for Scientific Research ``Demokritos'', 15310 Athens, Greece}
\affiliation{School of Applied Mathematical and Physical Sciences, National Technical University of Athens, 15780 Athens, Greece}
\author{J. Hizanidis}
\affiliation{Institute of Nanoscience and Nanotechnology, National Center for Scientific Research ``Demokritos'', 15310 Athens, Greece}
\affiliation{Crete Center for Quantum Complexity and Nanotechnology, Department of Physics, University of Crete, 71003 Heraklion, Greece}
\affiliation{National University of Science and Technology MISiS, Leninsky prosp. 4, Moscow, 
 119049, Russia}
\author{A. Provata}
\affiliation{Institute of Nanoscience and Nanotechnology, National Center for Scientific Research ``Demokritos'', 15310 Athens, Greece}

\date{Received: \today / Revised version: date}
% The correct dates will be entered by publisher
%
\begin{abstract}{

\quad We study the stable spatiotemporal patterns that arise in a 3D network of neuron oscillators, whose dynamics is described by the Leaky Integrate-and-Fire (LIF) model. More specifically, we 
investigate the form of the chimera states induced by a 3D coupling matrix with nonlocal topology. The observed patterns are in many cases direct 
generalizations of the corresponding 2D patterns, e.~g. spheres, layers and cylinder grids. We also find cylindrical and ``cross-layered'' chimeras that do not have
an equivalent in 2D systems. Quantitative measures are calculated, such as the ratio of 
synchronized and unsynchronized neurons as a function of the coupling range, the mean phase velocities and the distribution of neurons in mean phase velocities.
Based on these measures the chimeras are categorized in two families. The first family of patterns is observed for weaker coupling and 
exhibits higher mean phase velocities for the unsynchronized areas of the network. The opposite holds for the second family where the unsynchronized 
areas have lower mean phase velocities. The various measures demonstrate discontinuities, indicating criticality as the parameters cross from the first 
family of patterns to the second.

}
\end{abstract}

%

%\pacs{ 05.45.Xt, 89.75.-k, 87.19.lj

% Synchronization, nonlinear dynamics, 05.45.Xt
% Complex systems, 89.75.-k
% neuronal network dynamics, 87.19.lj
%} % end of PACS codes
%      %end of abstract
%
\maketitle
\section{Introduction}
\label{sec:intro}
Chimeras are collective states where synchronized and unsynchronized
regions coexist and were first identified in networks of non-locally coupled identical 
Kuramoto oscillators \cite{First_kuramoto2002, Name_abrams2004}.
Since then, the number of works dealing with this phenomenon
has grown immensely extending beyond phase oscillators and non-local coupling schemes (see \cite{Rev_Panaggio2015} and references within).
Experimentally, chimeras have been found in networks of coupled chemical \cite{ChemOsc_Tinsley2012},
optical \cite{Exp_Hagerstrom2012},
electrochemical \cite{Exp_schmidt2014,Exp_wickr2013} and mechanical oscillators \cite{MechOsc_Martens2013}. Further experimental evidence of the existence of chimera states can be found in electronic circuits with neuron-like spiking dynamics and first neighbor connections \cite{Exp_Gambuzza2014} and coupled pendula \cite{Exp_Kapitaniak2014}.
Other systems exhibiting chimeras are power grid networks \cite{powergrid1,powergrid2,powergrid3,powergrid4}, heart tissue during ventricular fibrillation
\cite{heart1,heart2,heart3,heart4} and superconducting metamaterials~\cite{SQUID_Lazaridis2015}. 

In the context of neuronal systems, it has been suggested that chimera states may be related to the
phenomenon of unihemispheric sleep~\cite{Theory_unihemispheric_Rattenborg2000} and epileptic seizures~\cite{Theory_epilepsy_Mormann2000,Theory_epilepsy_Mormann2003a,Theory_epilepsy_Andrzejak2016}. So far the works on neuronal networks are theoretical and involve 
various local dynamics  and network connectivities. 
More specifically,  there are works on 1D networks of Hodgkin-Huxley
\cite{Theory_HodgkinHuxley_Sakaguchi2006a, Theory_HodgkinHuxley_glaze2016}, FitzHugh-Nagumo \cite{Theory_2D_Omelchenko2012,Theory_fitz_omelchenko2015}, Hindmarsh-Rose \cite{Theory_Hizanidis2013, Theory_Bera2016} and Integrate-and-Fire neurons~\cite{Theory_Olmi2010}.
Although the majority of works refer to symmetrical coupling schemes, there are significant contributions
where it has been shown that hierarchical connectivities~\cite{Theory_fitz_omelchenko2015,Theory_Tsigkri2016,Sawicki2017}  
and modular networks~\cite{Theory_Celeg_Hizanidis2016} may also support chimera states. 

The 2D case was first studied in networks of phase oscillators, see refs.   
\cite{Theory_2D_Omelchenko2012,Theory_2D_Xie2015}. Recently,
 the Leaky Integrate-and-Fire (LIF) and FitzHugh Nagumo
models were employed and it was shown that both systems support 
similar 2D chimera patterns\cite{Theory_2D_Kasimatis2017}. The 3D problem has been discussed only by Maistrenko \emph{et al.} in \cite{Theory_3D_Maistrenko2015} and 
by Lau \emph{et al.} in \cite{Theory_3D_Lau2016}, where in both cases the Kuramoto model was used 
for the node dynamics. In \cite{Theory_3D_Maistrenko2015} spherical, cylindrical, crossed and layered chimera states were obtained, while in \cite{Theory_3D_Lau2016},
the existence of nontrivial ``linked'' and ``knotted'' chimera structures in 
oscillatory systems was discussed.

In the present manuscript we investigate the chimera states that arise in a 3D network of non-locally coupled LIF oscillators. The LIF model \cite{Theory_Tsigkri2016,LIF_Brunel2007} is an approximation of neuron dynamics and is reviewed in Sec.~\ref{sec:LIFmodel}.
In Sec.~\ref{sec:Steadychim} we present a detailed map of the obtained states 
in the relevant parameter space with focus on the stable chimera patterns.
The observed dynamics is discussed in detail and the different chimeras are sorted in categories.
In Sec.~\ref{sec:Quantitative analysis} we use quantitative measures to further validate the findings of Sec~\ref{sec:Steadychim} and critical behavior is discussed. We add an Appendix where tables relevant to Sec.~\ref{sec:Steadychim} are displayed.

\section{The Leaky Integrate-and-Fire model \& Quantitative measures}

In this section the LIF model is briefly recapitulated 
and the various assumptions and modifications we make for the 3D realization 
are discussed.
\label{sec:LIFmodel}

%\subsection{The LIF model for uncoupled neurons}
%\label{sec:1D}
\subsection{Single neuron dynamics}

The LIF model describes the dynamics of the neuron membrane potential as being similar to that of a capacitor with leakage current. This theoretical capacitor integrates for a finite period and discharges
when its potential reaches a threshold \(u_{th}\). 
As a result, it returns to the resting potential \(u_{rest}\), and the process of charging begins again. This model assimilates the basic dynamical features of spiking neurons and was first introduced by L. Lapique in 1907. The LIF cycle is represented by Eq. \ref{eq:lif01}, 
where Eq. \ref{eq:lif01}a describes the current integration and Eq. \ref{eq:lif01}b the discharge. In Eq. \ref{eq:lif01}
\(u\) is the potential of the neuron's
membrane and the term \(\mu\) represents the maximum possible $u$-value, because when \(u(t)=\mu\) the rate of potential 
change drops to 
zero (see Eq. \ref{eq:lif01}b). Note that \(u_{th} \leq \mu\) for consistent oscillatory motion.   

\begin{subequations} 
\label{eq:lif01} 
\begin{align} 
\frac{du(t)}{dt} =-u(t)+RI(t) \\ \lim_{\epsilon\rightarrow 0}u(t+\epsilon) \rightarrow u_{rest}\quad \text{when}\quad u\geq u_{th}.
\end{align}
\end{subequations}

%\subsection{Assumptions for uncoupled LIF elements}
%\label{sec:Assumptions coupling}

For limit intervals between \(u_{rest}\) and \(u_{th}\) the solution to Eq. (\ref{eq:lif01}) is:

\begin{align} \label{eq:lif01.1}  
u(t)=\mu -\left(\mu-u_{rest}\right)e^{-t}.
\end{align}

The period of the single LIF neuron can be calculated as:
\begin{equation} \label{eq:lif01.2} 
T_s=\ln \left[\left(\mu -u_{rest}\right)/\left(\mu - u_{th}\right)\right].
\end{equation}
One more parameter is added to the model in order to take into account the refractory period of the neuron.
The refractory period is an interval of time after a discharge during which the neuron is unable to start
charging again. The neuron oscillator remains at rest for time \(p_{r}\) after each discharge and thus the total period of the oscillator is \(T=T_{s}+p_{r}\):
 
\begin{equation} \label{eq:lif02}
u(t)=u_{rest} \quad \forall \text{t}: \medskip (\nu+1) T-p_{r}\leq t \leq (\nu+1)T,
\end{equation}
where \(\nu=0,1,2,\dots\) is the number of firings. This assumption completes the description of the dynamics of a single, isolated LIF neuron oscillator with refractory period. 

\subsection{3D coupling scheme}
%\label{sec:3D}

To introduce coupling in the system we consider a simple cubic lattice of size \(N\times N \times N\), where a neuron at lattice position (\(i,j,k\)) is coupled with a number of neighbors within a cube with edge \(2R+1\), where \(R\) is the coupling range. More specifically, the coupled neighbors with
coordinates (\(l,m,n\)) are inside a cubic area \(C_{R}(i,j,k)\) defined as:

\begin{equation}
\label{eq:lif03}
C_{R}(i,j,k)\equiv\{l,m,n\}\equiv
\begin{cases} 
      i-R \leq l \leq i+R\\ 
      j-R \leq m \leq j+R\\
      k-R \leq n \leq k+R.\\
   \end{cases}
\end{equation}

\begin{figure}[ht!]
\includegraphics[clip,width=1\linewidth,angle=0]{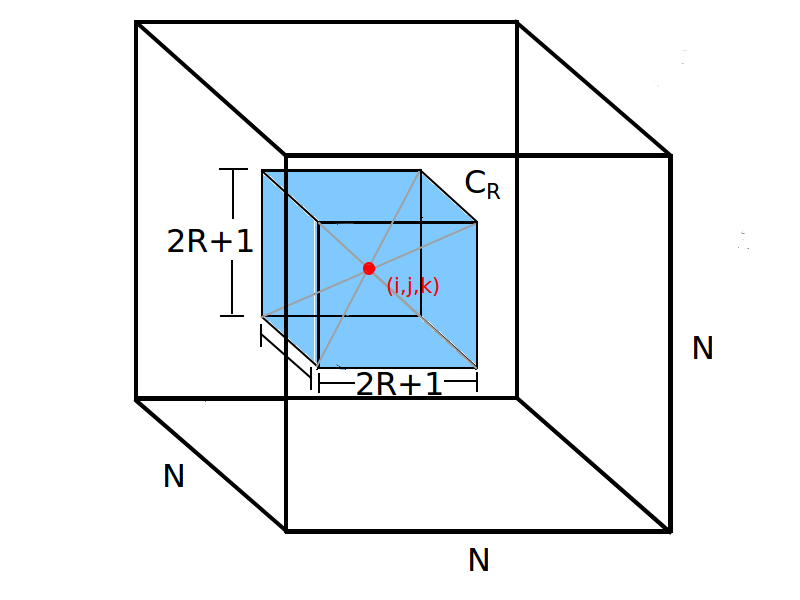}
\caption{
\label{3Dcoupling} (Color online) Depiction of the three dimensional neural network and
the coupled area \(C_{R}\) around neuron \((i,j,k)\).
}
\end{figure}

The number of neurons inside the coupling area \(C_{R}\) is given by \((2R+1)^{3}\) and the number of neurons coupled to the central 
(\(i,j,k\)) neuron is \((2R+1)^{3}-1\) (we exclude self-coupling).  We assume that the interaction between neuron (\(i,j,k\)) and its coupled neighbors is 
proportional to \(u_{lmn}(t)-u_{ijk}(t)\) with constant intensity \(\sigma_{ijk,lmn}\),  where (\(i,j,k\)) are constant and (\(l,m,n\)) run through the group of \((2R+1)^{3}-1\) 
coupled neurons inside \(C_{R}\) and holds for every neuron (\(i,j,k\)) in the network. The evolution of the potential at position \((i,j,k)\) is given by:
\begin{equation}\begin{split} \label{eq:lif04} 
\frac{du_{ijk}(t)}{dt}=-u_{ijk}(t)+\mu \\ 
+\frac{\sigma_{ijk,lmn}}{(2R+1)^{3}-1}\sum_{lmn \in C_{R}}\left[u_{lmn}(t) -u_{ijk}(t)\right].\end{split}\end{equation}

In this study the dimensions of the 3D lattice are \(N^{3}=27^{3}=19683\) neurons. We consider that all connections inside \(C_{R}\) around node \(i,j,k\) have the same coupling 
strength, independently of the position or the coupled neurons as follows:
\begin{equation}
\label{eq:sigma}
   \sigma_{ijk,lmn}=
   \begin{cases} 
      \sigma, & l,m,n \in C_{R}\\ 
      0, & \text{otherwise}. 
   \end{cases}
\end{equation}

Concerning the other parameters, \(\mu\) takes a constant value, \(\mu=1\) and without loss of generality the resting potential is \(u_{rest}=0\).
The refractory period is set to be \(p_{r}=0.21T_{s}\). This value is chosen to be comparable with refractory periods observed in natural 
neurons. The control parameters we use are the coupling constant \(\sigma\), the coupling range \(R\) and we also define the ratio of coupled neurons
as \(N_{R}=((2R+1)^{3}-1)/N^3\).
The initial conditions are pseudorandom and the boundary conditions are periodic in all three dimensions.

\subsection{Quantitative measures}
%\label{sec:qm}

A crucial quantitative measure in chimera state identification is the mean phase velocity. 
The calculation of mean phase velocities allows to identify synchronized and
unsynchronized areas of the system. To calculate the mean phase velocity we use data that correspond to points of time after the system has
reached a stable state. Mean phase velocities are calculated according to:
\begin{equation}
\label{eq:wmega}
   \omega_{ijk}(t)=2\pi \frac{C_{d_{ijk}}(t)}{t-t_{0}},
\end{equation}
where \(C_{d_{ijk}}(t)\) represents the total number of discharges the neuron \((i,j,k)\) had in the time interval \(t-t_{0}\) and \(t_{0}\) is the point in 
time after which \(\omega\) is calculated.

Another quantitative measure we use is the population of synchronized (and unsynchronized)
neurons. The neurons are characterized as synchronized or unsynchronized according to the 
following rule: The mean phase velocity range \(\Delta\omega_{max}\) is split into 100 segments
\(df=1\%\Delta\omega_{max}\), where \(df\) is the width of each segment. For each neuron (\(i,j,k\)) in the network we calculate the 
average of the mean phase velocity differences between neuron (\(i,j,k\)) and its first and second neighbors in the lattice and
as such the summation in Eq. (\ref{eq:synch_criterion}) runs through the first and
second neighbors of each element: 

\begin{equation}
\label{eq:synch_criterion}
\Delta\omega_{mean}(i,j,k)=\frac{\Sigma_{lmn}|\omega_{ijk}-\omega_{lmn}|}{26},
\end{equation}
where \(\omega_{lmn}\) is the mean phase velocity of a first or second neighbor of
neuron (\(i,j,k\)) and \(26\) is the number of first and second neighbors in a simple cubic lattice. If \(\Delta\omega_{mean} (i,j,k)\leq  3\%\Delta\omega_{max}\) then neuron (\(i,j,k\)) is considered 
synchronized otherwise it is considered unsynchronized.
The segmentation of the mean phase velocity range \(\Delta\omega_{max}\) is used
in Sec.~\ref{sec:Quantitative analysis}, where 
distributions of neurons in the mean phase velocity spectrum are calculated.

\section{Stable chimera patterns}
\label{sec:Steadychim}

\begin{figure}[ht!]
\includegraphics[clip,width=1\linewidth,angle=0]{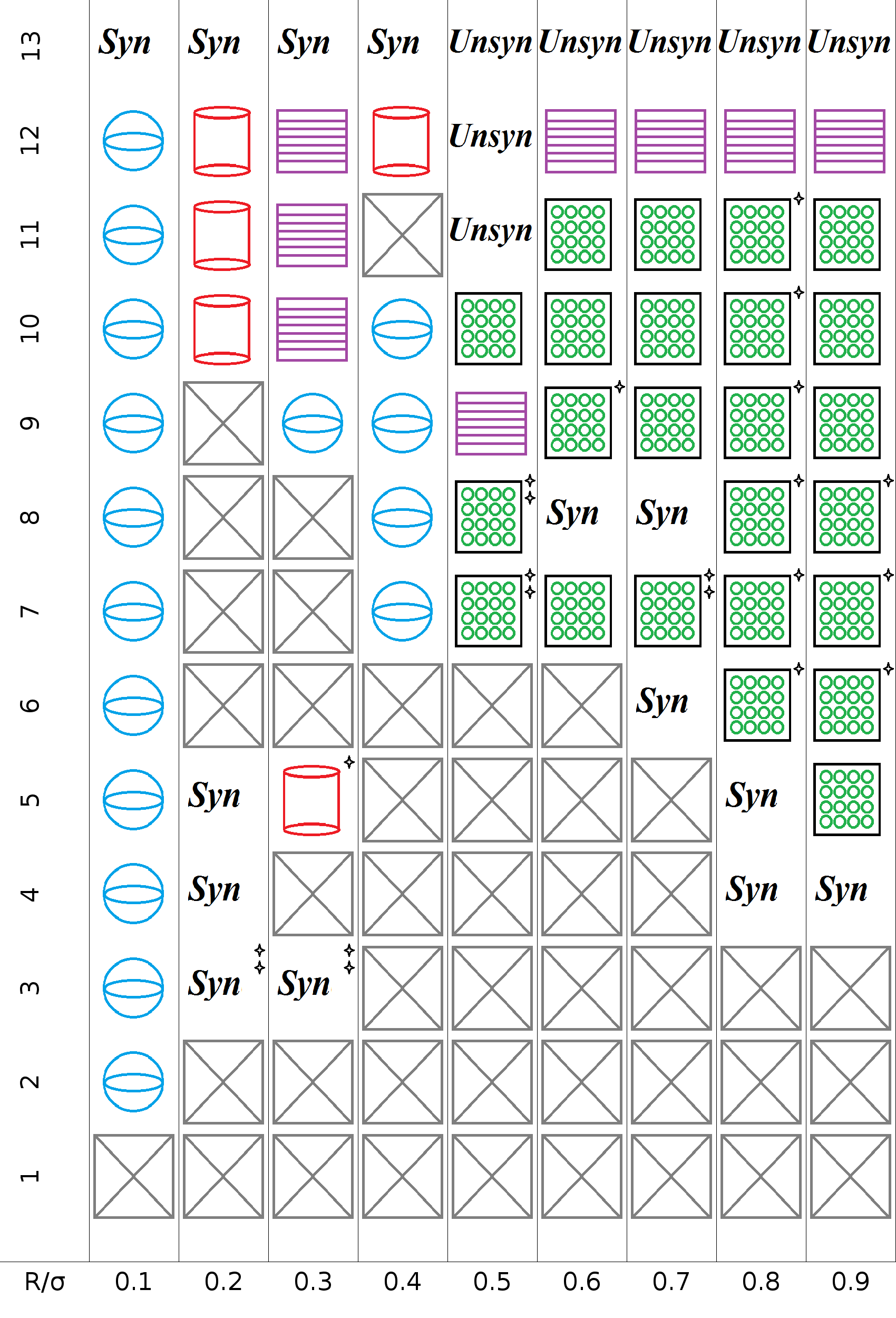}
%\caption{
%\label{map} (Color online) Coupling range versus coupling strength map of the different patterns identified in the oscillator network. }
\includegraphics[clip,width=1\linewidth,angle=0]{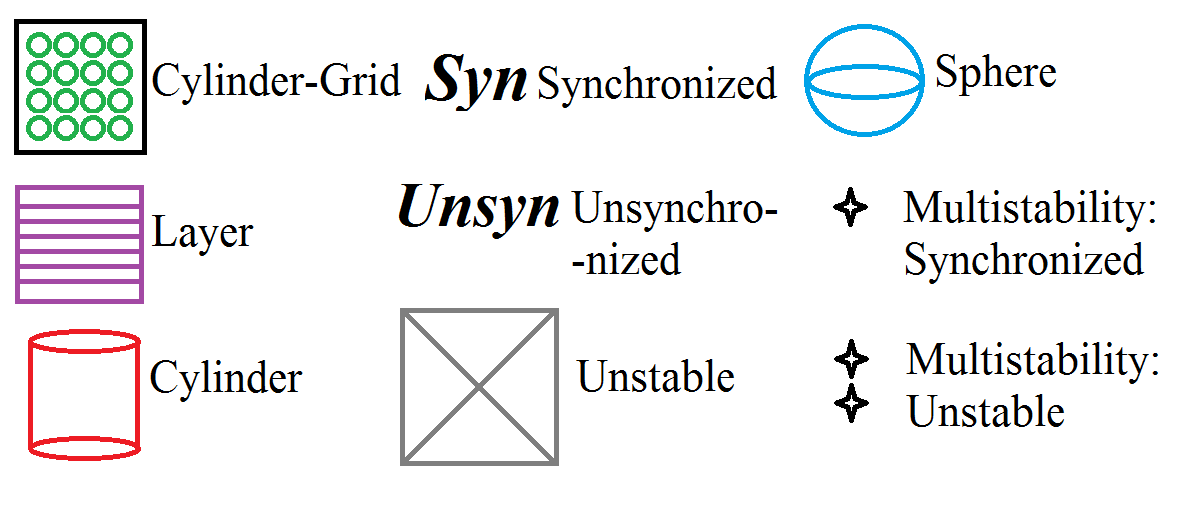}
\caption{
\label{map} (Color online)  Map of the different patterns identified in the 3D LIF oscillator network
in the  coupling range versus coupling strength plane.}
\end{figure}

In this section we present a variety of chimera patterns which arise as we vary the system parameters. We keep the refractory period to moderate values,
\(\Delta t_{ref}=0.21T_{s}\), compatible with the ones observed in natural neurons. The coupling constant takes values in the interval
\(0.1\leq \sigma \leq 0.9\)
and the coupling range \(R\) varies between \(1\leq R \leq 13\). Consequently the ratio of coupled neurons in the linked region is
\(N_{R}=((2R+1)^3-1)/N^3=> 0.14\% \leq N_{R} \leq 100\%\).
The system starts from random initial potentials that vary inside the interval
\(u_{rest} \leq u_{ijk}(t=0) \leq u_{th}\), \(i,j,k=1,...,N\).
Before getting into details, we present a collective map with the different patterns observed.
In Fig. \ref{map} cylindrical, spherical, layer and cylinder-grid chimeras are depicted. The unstable, synchronized and unsynchronized states are also noted.
The annotation is mostly self-explanatory and can be seen in the caption of Fig. \ref{map}. 

The inspection of Fig.~\ref{map} leads to the following general observations:
As the coupling range \(R\) increases, we notice that the unstable states gradually disappear.
In the limit of all-to-all connectivity chimera patterns are not possible since the system is either fully synchronized or unsynchronized.
For weak coupling the spherical chimera pattern is dominant while the increase of the coupling strength gives rise to the cylinder-grid
pattern or total synchronization depending on the initial conditions. Another interesting effect is that for \(\sigma\geq0.5\) the system may stabilize in 
different patterns, depending on the initial conditions.
This observation alongside others that we discuss in the following sections leads us to hypothesize that a critical point of the system exists between 
\(\sigma=0.4-0.5\), above which the system exhibits multistability. Also, almost all multistable states are cylinder-grid states with the exception of single cylinder state (\(\sigma=0.3, R=5\)).
In the next three subsections we present in detail each pattern.

\subsection{Spheres}
\label{sec:Spheres}

\begin{figure}[ht!]
\includegraphics[clip,width=1\linewidth,angle=0]{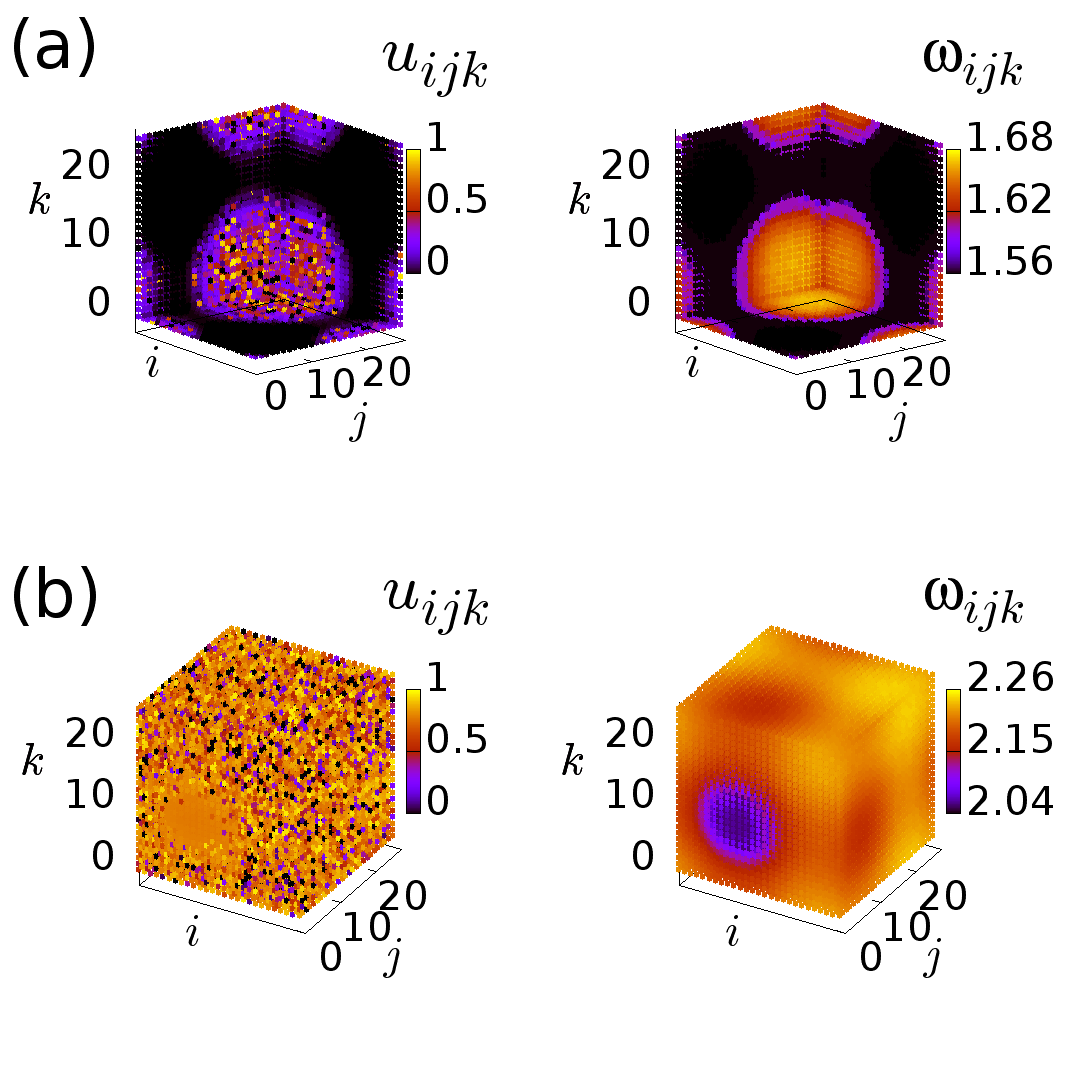}
\caption{
\label{fig1} (Color online) Spherical Chimeras: Each elementary cube inside the 3D figure
represents a neuron and the color of the cube represents either the membrane potential \(u_{ijk}\) of the neuron (left) or the mean phase velocity 
\(\omega_{ijk}\) (right).
Parameters: (a) \(R\)= 8, \(N_{R}\)= 24.96\%, \(\sigma\)= 0.1 
and (b) \(R\)= 10, \(N_{R}\)= 47.05\% , \(\sigma\)= 0.4.
Other parameters: \(\mu=1\), \(u_{th}=0.98\), \(u_{rest}=0\), \(p_{r}=0.21T_{s}\).
Periodic boundary conditions are applied and initial conditions are randomly selected.
}
\end{figure}

One of the most common patterns for low coupling strength is the sphere pattern. This pattern is a direct 3D generalization of the 2D spot chimera and
the 1D single chimera pattern. These chimeras consist of two regions, one spherical and a surrounding
region. The spherical region is convex while its supplementary surrounding region is concave. For \(\sigma\)=0.1 the spherical region is unsynchronized
and has greater mean phase velocities than the surrounding synchronized region (Fig. \ref{fig1}a).
The opposite holds for \(\sigma\)=0.3 and 0.4 where the synchronized domain is inside the spherical region and the unsynchronized in the
surrounding region (Fig. \ref{fig1}b). The mean phase velocities of the surrounding area are greater than those inside the sphere. In both cases the 
synchronized regions have lower mean phase velocities than the unsynchronized. The spherical chimera for \(\sigma=0.3\) (not shown) comprises four 
spherical regions which are formed near the periodic boundaries and overlap with each other resulting in a pattern very similar to the cross chimera 
reported in Y. Maistrenko \emph{et al.}\cite{Theory_3D_Maistrenko2015} although less symmetric.  

Spherical chimeras are observed for the following control parameter regions: [\(\sigma\)= 0.1 \& \(2 \leq R \leq 12\)], [\(\sigma\)= 0.3 \& \(R\)=9], [\(\sigma\)= 0.4 \& \(7 \leq R \leq 10\)].

\subsection{Cylinders}
\label{sec:Cylinders}

\begin{figure}[ht!]
\includegraphics[clip,width=1\linewidth,angle=0]{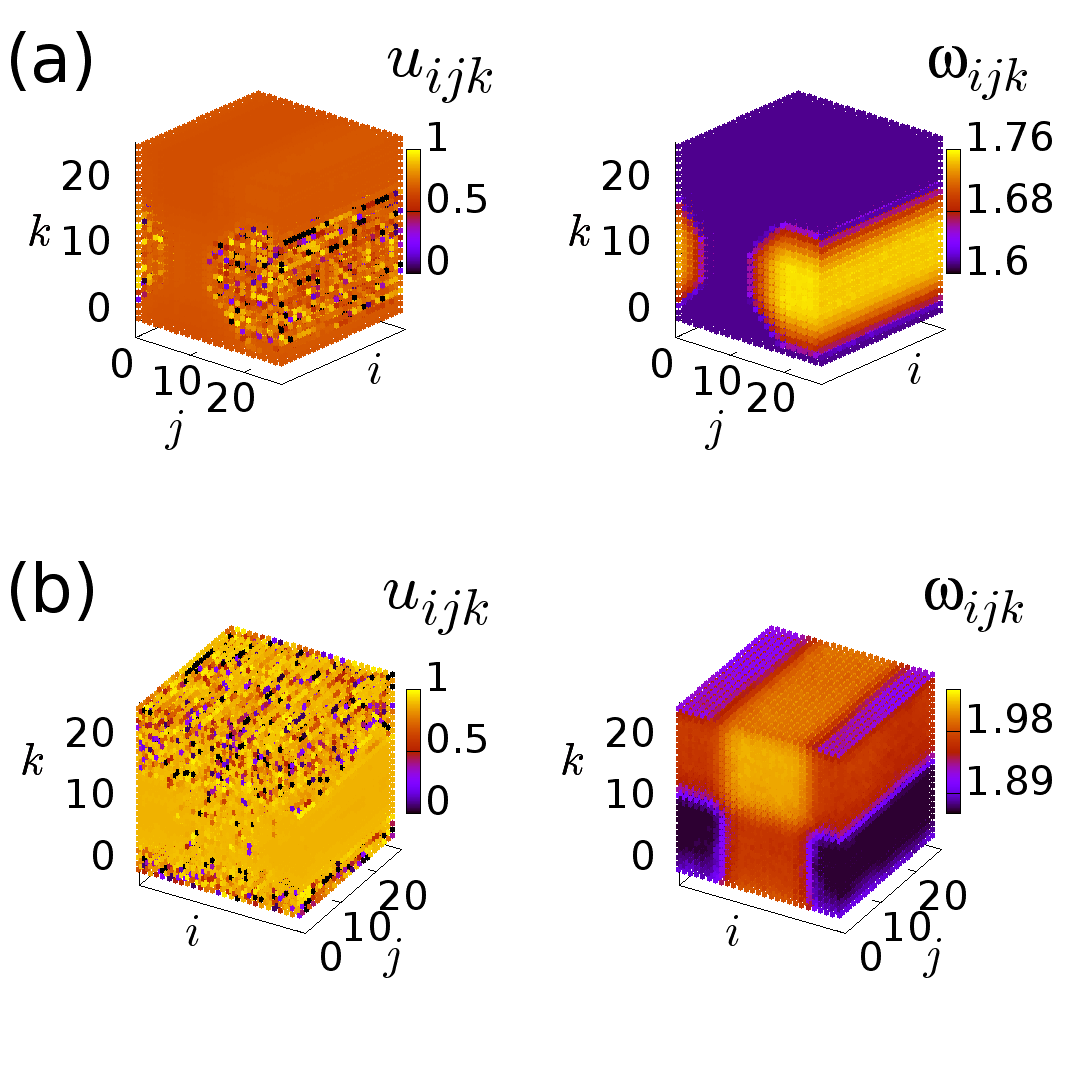}
\caption{\label{fig2} (Color online) Cylindrical Chimeras: The membrane potential 
\(u_{ijk}\) (left) and the mean phase velocities \(\omega_{ijk}\) (right) are displayed. a) the depicted cylinder is unsynchronized 
b) the surrounding area is unsynchronized and the cylinder is synchronized. Parameters: (a) \(R\)= 11, \(N_{R}\)= 61.81\% , \(\sigma\)= 0.2
and (b)  \(R\)= 12, \(N_{R}\)= 79.38\% , \(\sigma\)= 0.4. Other parameters as in Fig. \ref{fig1}.
}
\end{figure}

For mostly large values of \(R\) and small to medium values of \(\sigma\) we observe the formation of stable cylindrical patterns.
In this type of chimera the 
network splits in two regions, the cylindrical region and the surrounding area. In this case, contrary to the 
spherical chimera (Sec. \ref{sec:Spheres}),
the cylinder runs through the length of the network and reaches its periodic boundaries.
In conjunction with Sec. \ref{sec:Spheres}, for low values of \(\sigma\) the cylindrical region is unsynchronized and has greater mean phase velocities 
than the surrounding area (see Fig. \ref{fig2}a),  while the exact opposite is true for greater \(\sigma\) values (see Fig. \ref{fig2}b). In the latter case 
the unsynchronized region demonstrates larger mean phase velocities as was the case for the spherical patterns. In Fig. \ref{fig2}b, in the mean phase 
velocity profiles, the
unsynchronized area is split into mean phase velocity groups that exhibit very small mean phase velocity ranges and for that reason they can be 
mistaken for synchronized areas. In Sec. \ref{sec:Distribution} it is shown that indeed these areas comprise unsynchronized neurons.

It is important to point out that for \(\sigma\)=0.4 we have spherical chimeras for \(R\)=7,8,9 and 10 (see \ref{sec:Spheres}) 
and cylindrical chimeras for \(R\)= 12, while for R= 11 the system is unstable. Therefore, by increasing the coupling radius \(R\), the spherical patterns are replaced first with an unstable state and then by 
cylindrical chimeras. For \(\sigma\)=0.2 and 0.3 the cylindrical area is unsynchronized and has 
greater mean phase velocities (similar to Fig. \ref{fig1}a) while for \(\sigma\)= 0.4 the 
opposite is true (Fig. \ref{fig1}b). This is consistent with what was observed for the spherical 
chimeras in \ref{sec:Spheres}.

\subsection{Layers}
\label{sec:Layers}

\begin{figure}[ht!]
\includegraphics[clip,width=1\linewidth,angle=0]{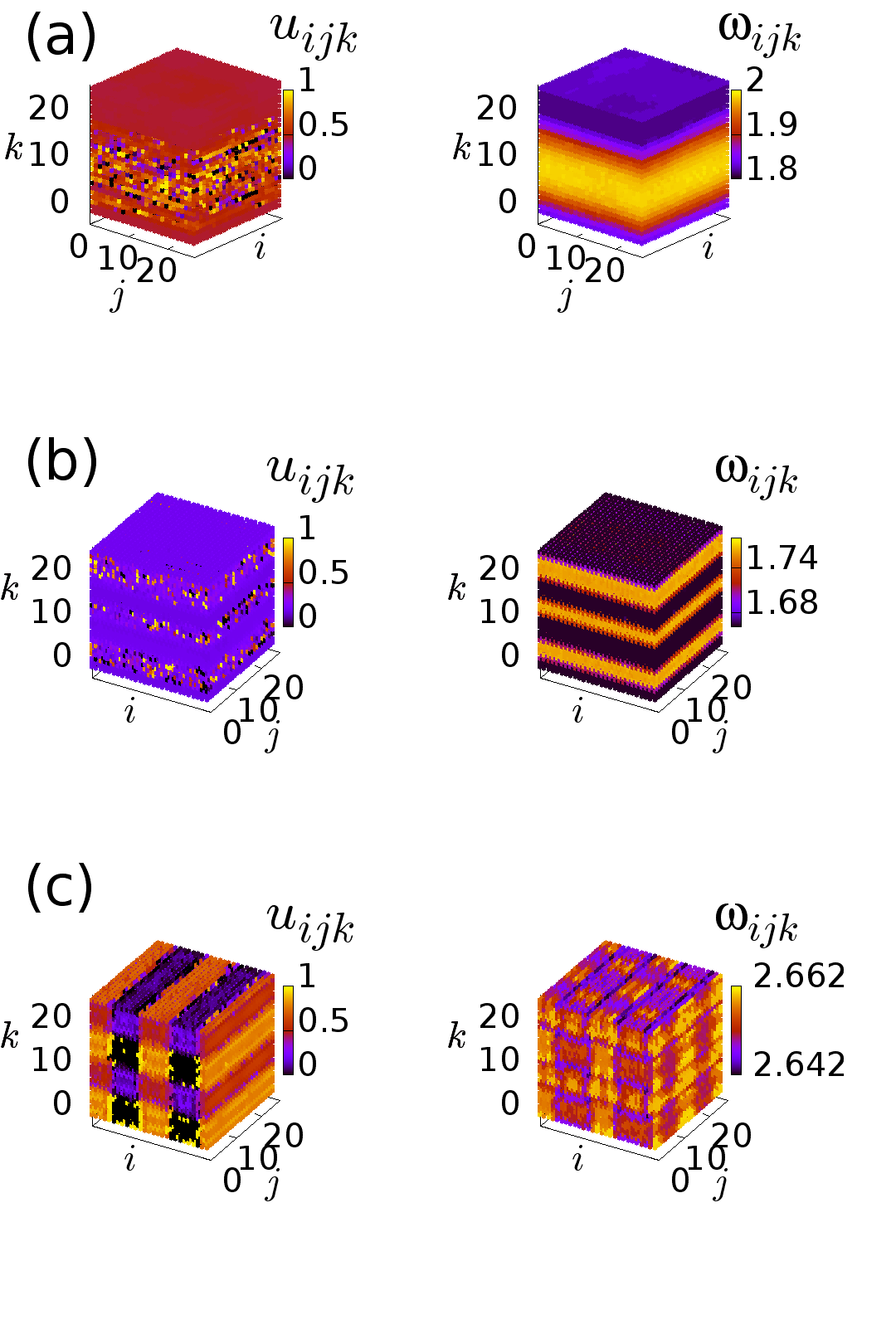}
\vspace{-1.5cm}
\caption{\label{fig3} (Color online) Layer Chimeras: The membrane potential (left) and mean 
phase velocities (right) of each neuron inside the network. a) an unsynchronized layer is formed in the middle and 
a synchronized region forms around it; b) three synchronized and three unsynchronized regions alternate in space; c) multiple vertical and horizontal layers cross the system. The  cross-layered
pattern has very limited unsynchronized areas while the mean phase velocity range is one order of magnitude smaller than (a) and (b).  Parameters: (a) \(R\)= 11, \(N_{R}\)= 61.81\%,
\(\sigma\)=0.3, (b) \(R\)= 12, \(N_{R}\)=79.38\% , \(\sigma\)=0.3 and (c) \(R\)= 12, \(N_{R}\)=79.38\% , \(\sigma\)=0.6.
Other parameters as in Fig. \ref{fig1}.
}
\end{figure}

The layered pattern is observed mostly for large values of the coupling range \(R\). In the layer chimera regimes one or multiple layers run through the network 
of coupled neurons. The alternating areas formed differ in that they may be synchronized or unsynchronized and even when they exhibit synchronization 
they do so in different mean phase velocities. Examples of both single and multilayer configurations can be seen in Figs. \ref{fig3}a,b. 

Layer patterns were observed for large values of \(R\) (\(R=9,10,11,12\)).
For \(\sigma\)=0.3 \& \(R\)=12 we 
observed multiple layer chimeras (see Fig. \ref{fig3}b) while for R=12 and \(\sigma\)=0.6, 0.7, 0.8, 0.9 we observe cross-layered (see Fig. \ref{fig3}c).
The rest (\(\sigma\)=0.3 \& \(R\)=(10,11), \(\sigma\)=0.5 \& \(R\)=9) were single layer chimeras like the one shown in Fig. \ref{fig3}a.
As a result, there seems to exist a relation between the multiplicity of the layers and 
the coupling range, with the multiplicity increasing with coupling range \(R\). Also greater values of the coupling constant give rise to cross-layered 
chimeras. 

It is important to mention that many final states which were considered synchronized, and thus are not discussed here, were actually layer 
patterns but with very small difference between maximum and minimum mean phase velocities.
While in the chimera patterns (such as those displayed in Fig. \ref{fig3}) this difference 
$\Delta \omega$ is of the order of \(10^{-1} - 10^{-2}\), in the states considered as synchronized $\Delta \omega$ was of the order
of \(10^{-4}\) or even smaller. In addition, many unstable states were observed where layers were formed at some point but did not keep the same orientation, multiplicity
or form in time. The patterns in Fig. \ref{fig3} a and b are 3D generalizations of the stripe pattern observed in 2D networks 
\cite{Theory_2D_Kasimatis2017}, while there is no equivalent in 1D chimeras. The cross-layered pattern (Fig. \ref{fig3}c) differs from the rest in
that the unsynchronized region has lower mean phase velocities than the various groups of synchronized neurons. The only other pattern that exhibits 
lower mean phase velocities in the unsynchronized regions is the cylinder-grid pattern which is discussed in the next section,
Sec. \ref{sec:Cylinder-Grid}.

\subsection{Cylinder-Grid patterns}
\label{sec:Cylinder-Grid}

\begin{figure}[ht!]
\includegraphics[clip,width=1\linewidth,angle=0]{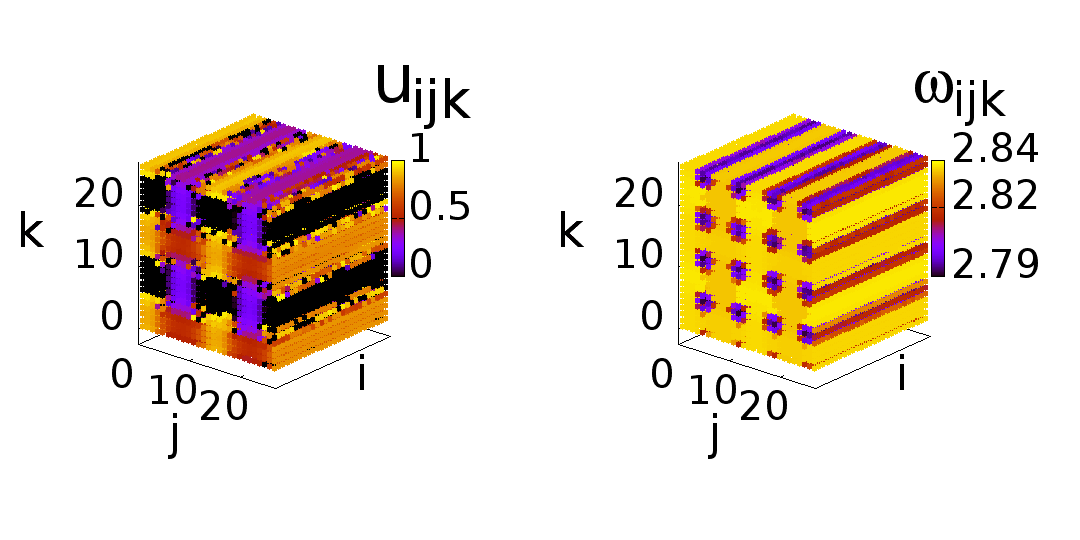}
\caption{\label{fig4} (Color online) Cylinder-grid chimeras: The membrane potential (left) and mean phase velocities (right) can be observed in this figure. Parameters: \(R\)= 10, \(N_{R}\)= 47.05\% , \(\sigma\)=0.7. Other parameters as in Fig. \ref{fig1}. }
\end{figure}

 The cylinder-grid pattern is mostly observed for larger values of the coupling constant \(\sigma\). We can see in Fig. \ref{fig4} that small cylindrical regions 
of unsynchronized neurons run through the network in such a way that a grid is formed. The multiplicity of the cylinders is 
usually $4 \times 4$. The unsynchronized cylinders have lower mean phase velocities than the
surrounding synchronized area. As for the synchronized areas they are arranged in compact groups of neurons which, although they have the same mean 
phase velocities, are locked in different phases. This means that the system exhibits phase grouping.
It is important to mention that this pattern is strongly influenced by the initial conditions. More specifically for different initial conditions we obtain either
cylinder-grid or fully synchronized patterns and less frequently unstable patterns. The same is not true for the rest of the patterns with very few exceptions. 

The cylinder-grid pattern is similar to the crossed  pattern (Fig. \ref{fig3}c); what makes them different is that in the cross-layered pattern the unsynchronized area is not organized in cylinders which can be seen clearly by comparing the mean phase velocities figures. Nevertheless we mention again that
grid-stripe and cross-layered patterns are the only ones where the unsynchronized areas have lower mean phase velocities. This phenomenon
was also observed in 1D LIF chimeras where for \(\sigma > 1\) the unsynchronized regions have lower \(\omega\)'s than the synchronized regions
\cite{Theory_Tsigkri2017}. Elaboration on this observation for the 3D LIF follows in Sec. \ref{sec:Distribution}

Cylinder-Grid patterns were observed for [\(\sigma\)=0.5 \& \(R\)=(7, 8, 10)], [\(\sigma\)=0.6 \& \(R\)=(7, 9, 10, 11)],  [\(\sigma\)=0.7 \& \(R\)=(7, 9, 
10, 11)],  [\(\sigma\)=0.8 \& \(R\)=(6 to 11)],  [\(\sigma\)=0.9 \& \(R\)=(5 to 11)]. It is apparent that large values of \(\sigma\) and \(R\) favor the appearance
of the cylinder-grid pattern. This pattern is a generalization of the grid pattern observed in 2D LIF networks shown in ref.
\cite{Theory_2D_Kasimatis2017}.

\section{Quantitative analysis of chimera pattern characteristics}
\label{sec:Quantitative analysis}

\subsection{Mean phase velocities}
\label{sec:Mean phase velocities}

\begin{figure}[ht!]
\includegraphics[clip,width=1\linewidth,angle=0]{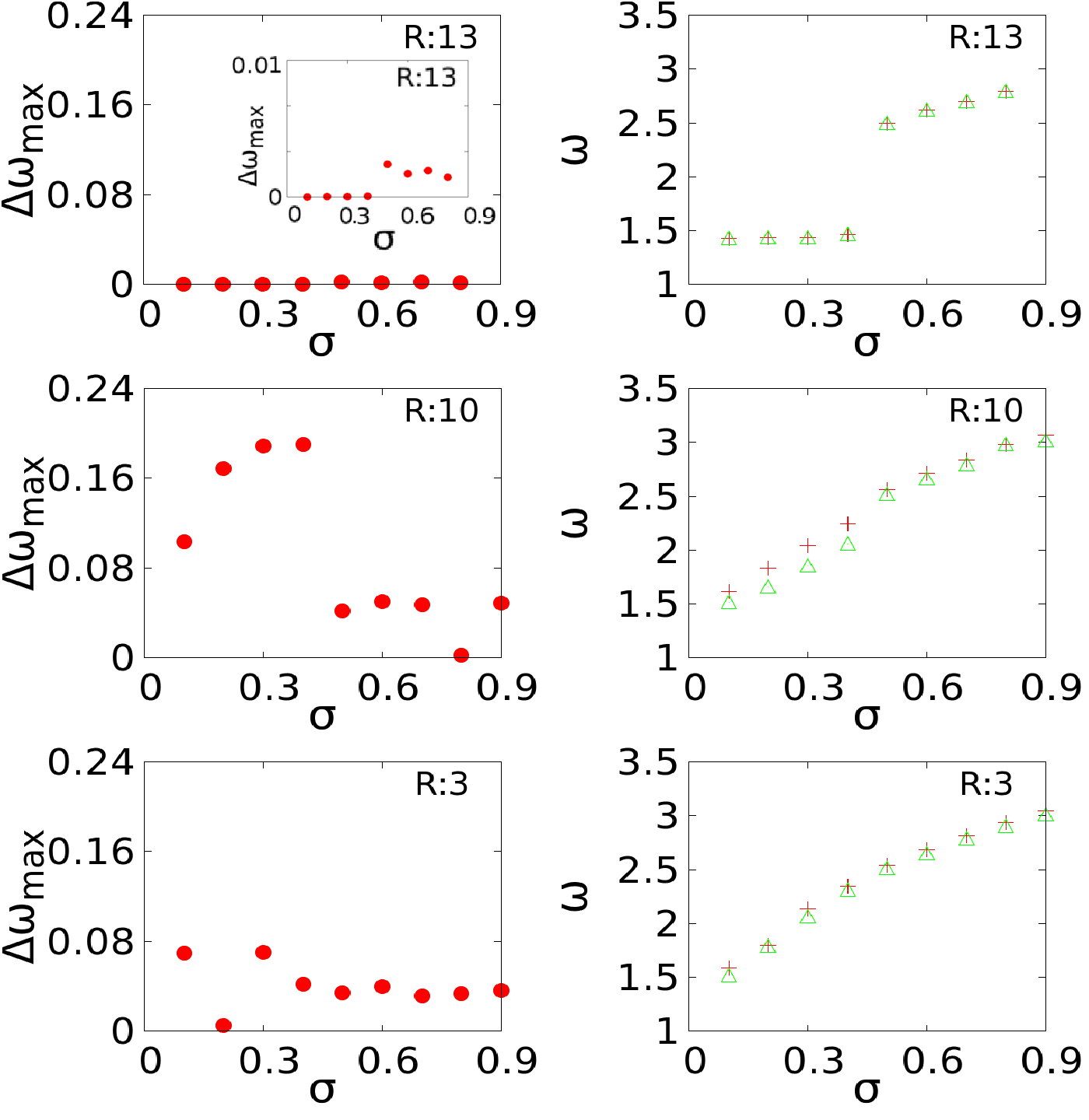}
\caption{\label{dw_wmaxmin_sample} (Color online)  \(\Delta\omega_{max}\) versus \(\sigma\) is depicted in the left pannels
while \(\omega_{max}\) (red-crosses) and \(\omega_{min}\) (green-triangles) are depicted in the right pannels. The top pannels correspond to all to all connectivity \(R=13\)
the middle pannels to \(R=10\) and bottom pannels to \(R=3\). Other parameters as in Fig. \ref{fig1}.}
\end{figure}

The maximum and minimum values of the mean phase velocities were extracted for each
parameter pair \(\sigma,R\). In Fig. \ref{dw_wmaxmin_sample} we present the \(\omega_{max}\), \(\omega_{min}\) and
\(\Delta\omega_{max}\) versus \(\sigma\) for \(R=3\) (small coupling range), \(R=10\) (intermediate coupling range)
and \(R=13\) (all to all coupling). The complete table  for \(R=1\) to \(R=13\) 
is available in the Appendix, Fig. \ref{dwA}. 

The first observation is that mean phase velocities increase almost linearly with the coupling constant. This is in line with what was observed for 
the 2D network in \cite{Theory_2D_Kasimatis2017} and it is intuitively expected. Since the increase of the membrane potential in a time unit is proportional 
to the coupling constant it is expected that higher \(\sigma\) values would result in higher mean phase velocities.
The second and more important observation made is that for \(R=7\) to \(R=12\) the linearity of the aforementioned increase breaks between 
\(\sigma=0.4\) and \(\sigma=0.5\) and two different linear segments emerge.
Exactly at these \(\sigma\) values we have change of the chimera states from one pattern to another and while on the ``left" side of the \(\sigma\) \(=0.4-0.5\) 
interval different chimera patterns are recorded, on the ``right" side we have mostly the same cylinder-grid pattern, as can be seen in the parameter map
Fig. \ref{map}. This change is also apparent in the \(\Delta\omega_{max}\) diagram since 
for $\sigma=0.1-0.4$ the mean phase
velocity range is significantly larger than in the $\sigma=0.5-0.9$ interval. Equivalent linearity gaps appear for smaller values of \(R\) 
but not always for the same coupling constant and the gap between the two linear segments gets smaller as the coupling range decreases
(see Fig. \ref{dw_wmaxmin_sample}). The linearity break intensifies for all to all connectivity \(R=13\) while the \(\Delta\omega_{max}\) also splits in 
two segments (in conjuction to the \(R=7\)-\(12\) case) where the lower \(\sigma\) segment is synchronized (\(\Delta\omega_{max} = 0\))
and the greater \(\sigma\) segment has very low but non zero \(\Delta\omega_{max}\) and thus is considered unsynchronized.   

\subsection{Distribution of neurons in mean phase velocity intervals}
\label{sec:Distribution}

\begin{figure}[ht!]
\includegraphics[clip,width=1.1\linewidth,angle=0]{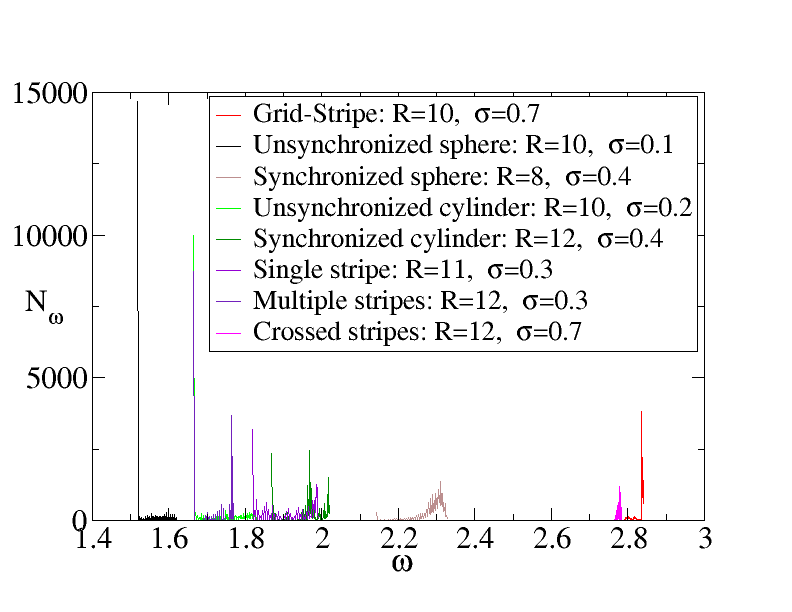}
\caption{\label{omega_distr}(Color online) \(N_{\omega}\) vs \(\omega\) for different chimera patterns. Identical chimera
patterns share the same distribution profile regardless of their control parameters. Other parameters as in Fig. \ref{fig1} .
}
\end{figure}

It is observed that each chimera pattern (spheres, cylinders, layers,
cylinder-grids) has a unique distribution profile. The term distribution profile refers to the shape of
\(N_{\omega}\) vs \(\omega\) diagram, where \(N_{\omega}\) is the number of neurons that have mean
phase velocities in the interval \(\omega \pm 1\%\Delta\omega_{max}\) (\(\Delta\omega_{max}=\)mean 
phase velocity range). In Fig. \ref{omega_distr} the various distributions are shown in the same 
diagram (one representative case for each chimera pattern). The 
important observation here is that most patterns share the same mean phase velocities in the
range $\omega \in [1.7-2]$ with 
the exceptions being the spherical chimeras (synchronized and unsynchronized), cylinder-grid 
chimeras and cross-layered chimeras for $\omega>2$.

\begin{figure*}[ht!]
\includegraphics[clip,width=1.05\linewidth,height=0.3857\linewidth,angle=0]{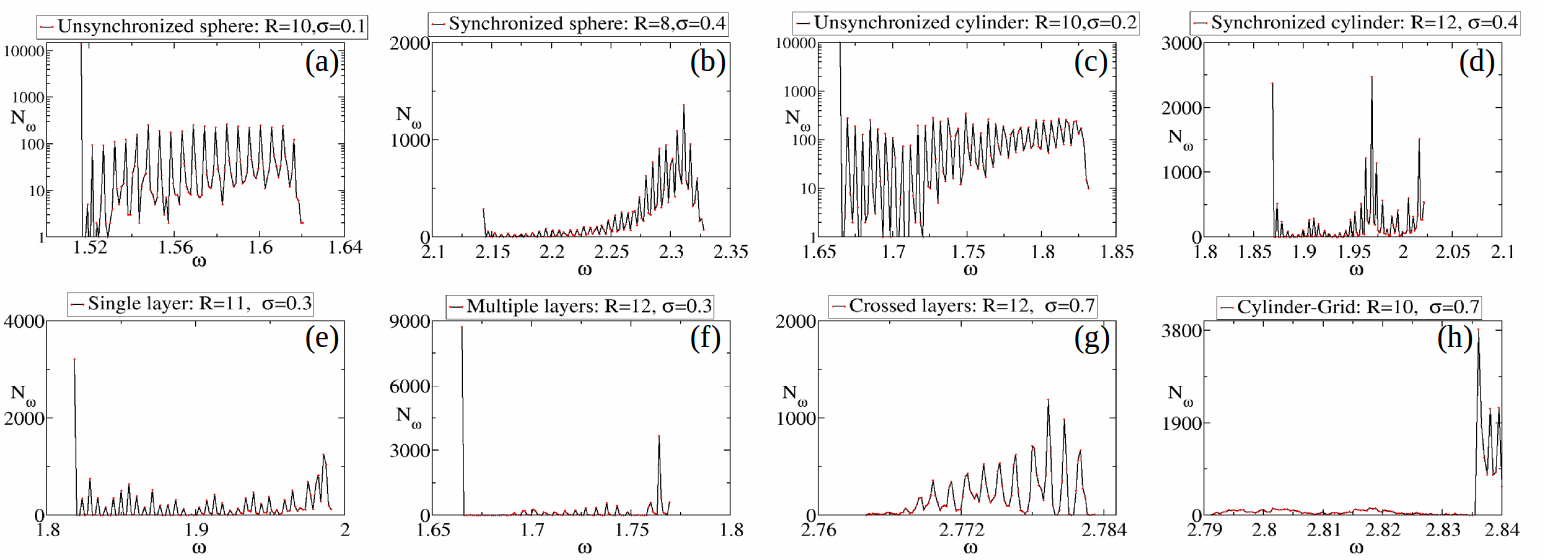}
\caption{\label{multi} (Color online) Distribution profile for each stable chimera state. For clarity, the distributions for the unsynchronized sphere and
unsynchronized cylinder are depicted on a logarithmic y-axis, while the rest have linear axes. Other parameters as in Fig.~\ref{fig1}.}
\end{figure*}

Cylinder-grid chimeras and cross-layered chimeras have many similarities, especially
in the membrane potential diagrams (see Fig. \ref{fig4} and Fig. \ref{fig3}c) and we see now 
that their distributions are relatively close (high mean phase velocities), but they indeed are different patterns since their distribution profile is quite different despite them being in 
adjacent mean phase velocity intervals. This will become clear as we investigate each distribution separately in Fig. \ref{multi}.
As we go through the distribution profiles keep in mind that in order to better depict them the axes of each 
pattern in Fig. \ref{multi} are different. In several cases the y-axis is drawn in logarithmic scale for clarity. To compare the distribution profiles  Fig. 
\ref{omega_distr} should be used.

%\begin{figure}[ht!]
%\includegraphics[clip,width=1\linewidth,angle=0]{./distr_sd2_R10_s01_inc_sphere.png}
%\caption{\label{inc_sphere} (Color online) Distribution profile for unsynchronized spheres:
%\(R=10\), \(\sigma=0.1\). Other parameters as in Fig. \ref{fig1}}
%\end{figure} 

The unsynchronized sphere distribution (Fig. \ref{multi}a)
 has one high peak in low mean phase velocities and 19
considerably lower secondary peaks dispersed in the mean phase velocity spectrum. The primary
peak corresponds to the synchronized area around the sphere, while the smaller peaks correspond
to the various mean phase velocities that neurons have inside the sphere. Contrary to that,
for the synchronized sphere (Fig. \ref{multi}b) the primary peak is greatly diminished and now
corresponds to the inside of the sphere. The secondary peaks correspond to the various mean phase velocities
the neurons exhibit in the surrounding area. 

%\begin{figure}[ht!]
%\includegraphics[clip,width=1\linewidth,angle=0]{./distr_sd2_R8_s04_coh_sphere.png}
%\caption{\label{coh_sphere} (Color online) Distribution profile for synchronized spheres:
%\(R=8\), \(\sigma=0.4\). Other parameters as in Fig. \ref{fig1}}
%\end{figure}

For the unsynchronized cylinder we can see in Fig. \ref{multi}c that the distribution
profile is similar to the unsynchronized sphere (keep in mind that both distributions are depicted for a logarithmic y-axis)
The difference between them is that more neurons belong to the unsynchronized area making the primary peak considerably lower and there are 31 secondary 
peaks instead of 19. It should be mentioned that for both spheres and cylinders the number of
peaks may vary slightly (increases or decreases by 1 to 5 peaks) depending on the parameters.

%\begin{figure}[ht!]
%\includegraphics[clip,width=1\linewidth,angle=0]{./distr_sd3_R10_s02_inc_cylinder.png}
%\caption{\label{inc_cylinder} (Color online) Distribution profile for unsynchronized cylinders:
%\(R=10\), \(\sigma=0.2\). Other parameters as in Fig. \ref{fig1}}
%\end{figure}

The synchronized cylinder distribution profile has an average-height primary peak and two major secondary peaks with various satellite peaks around them (Fig. \ref{multi}d).
The synchronized cylinder corresponds to the primary peak, while the secondary peaks correspond to 
two major and other less populated satellite mean phase velocities. These satellite peaks 
prohibit the system from passing to a completely group-synchronized state and destabilize the two 
major secondary peaks enough to cause desynchronization.

%\begin{figure}[ht!]
%\includegraphics[clip,width=1\linewidth,angle=0]{./distr_sd2_R12_s04_coh_cylinder.png}
%\caption{\label{coh_cylinder} (Color online) Distribution profile for synchronized cylinders:
%\(R=12\), \(\sigma=0.4\). Other parameters as in Fig. \ref{fig1}}
%\end{figure}

In the single layer chimera state (Fig. \ref{multi}e) the system splits into two areas: the unsynchronized layer and a
synchronized area. The primary peak corresponds to the synchronized area, while the secondary
peaks make up the unsynchronized layer.
 
%\begin{figure}[ht!]
%\includegraphics[clip,width=1\linewidth,angle=0]{./distr_sd2_R11_s03_single_stripe.png}
%\caption{\label{single_layer} (Color online) Distribution profile for single layer:
%\(R=11\), \(\sigma=0.3\). Other parameters as in Fig. \ref{fig1} }
%\end{figure}

In the multiple layer chimera state (Fig. \ref{multi}f) 
there exist alternating synchronized areas with different mean phase velocities and between them form 
unsynchronized layers. In the corresponding plot
we observe the primary peak and one major secondary peak with no satellites. The primary peak
corresponds to the synchronized low mean phase velocity areas while the major secondary peak to those with high mean phase velocity. The rest of the 
secondary peaks make up the unsynchronized 
layers.

%\begin{figure}[ht!]
%\includegraphics[clip,width=1\linewidth,angle=0]{./distr_sd2_R12_s03_multi-stripe.png}
%\caption{\label{multi-layer} (Color online) Distribution profile for multiple layers:
%\(R=12\), \(\sigma=0.3\). Other parameters as in Fig. \ref{fig1}}
%\end{figure}

The cross-layered chimera state has a distribution profile  
where there is no primary peak (Fig. \ref{multi}g). Instead there are many major secondary peaks which represent various 
mean phase velocity groups in the system. At the same time there is a small number of 
unsynchronized neurons dispersed in other mean phase velocities.

%\begin{figure}[ht!]
%\includegraphics[clip,width=1\linewidth,angle=0]{./distr_sd2_R12_s07_cross_stripes.png}
%\caption{\label{cross_layers} (Color online) Distribution profile for crossed layers:
%\(R=12\), \(\sigma=0.7\). Other parameters as in Fig. \ref{fig1}}
%\end{figure}

%\begin{figure}[ht!]
%\includegraphics[clip,width=1\linewidth,angle=0]{./distr_sd3_R10_s07_grid-stripe.png}
%\caption{\label{Cylinder-Grid} (Color online) Distribution profile for Cylinder-Grid:
%\(R=10\), \(\sigma=0.7\). Other parameters as in Fig. \ref{fig1}}
%\end{figure}

In the cylinder-grid pattern (Fig. \ref{multi}h) we observe mean phase velocity grouping and a grid of parallel cylinders that 
runs through the system. There is no primary peak and we have three major secondary peaks that
consist the mean phase velocity groups. The rest of the profile distribution is a continuous
line that corresponds to the unsynchronized cylinders.

A general observation about mean phase velocity distributions is that as the coupling constant
\(\sigma\) increases the hight of the primary peak decreases and this gives rise to secondary major peaks.
For greater \(\sigma\) values the primary peak vanishes, which leads to cylinder-grid chimeras and
less frequently, to cross-layered chimera states. As a result the unsynchronized areas in cylinder-grid
and cross-layered chimeras exhibit lower mean phase velocities than the synchronized areas. The opposite is
true for the rest of chimera states observed.

\subsection{Synchronized and Unsynchronized populations}
\label{sec:Synchronized and Unsynchronized populations}
\begin{figure}[ht!]
\includegraphics[clip,width=1\linewidth,angle=0]{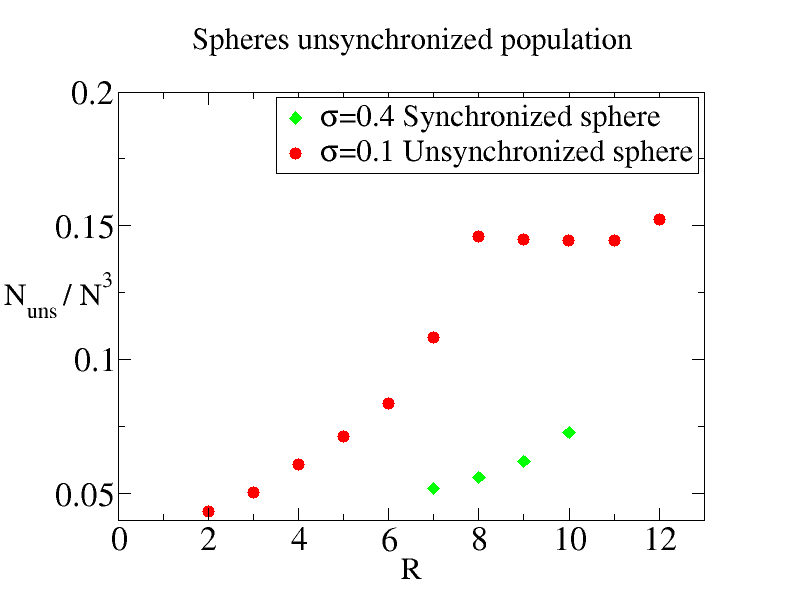}
\caption{\label{sphere-inc-pop} (Color online) Relative number of unsynchronized neurons for different spherical chimera states. Other parameters as in Fig.~\ref{fig1}.}
\end{figure}

The synchronized and unsynchronized relative populations were determined as described
in  Sec. \ref{sec:LIFmodel}. There are two general observations about the percentage
of unsynchronized neurons. First in the spherical chimera states we observe an exponential 
increase of unsynchronized neurons for \(R=2\) to \(R=8\) and \(\sigma=0.1\) as seen in
Fig. \ref{sphere-inc-pop} (red circles). After \(R=8\) the percentage of unsynchronized
neurons is relatively stable.

For \(R=7\) to \(R=10\) and \(\sigma=0.4\) we have synchronized
spheres and the rest of the network is unsynchronized. In this case the unsynchronized neurons
increase exponentially without reaching any upper threshold (green diamonds). For values of \(R >\)10 this pattern is
not observed.

\begin{figure}[ht!]
\includegraphics[clip,width=1\linewidth,angle=0]{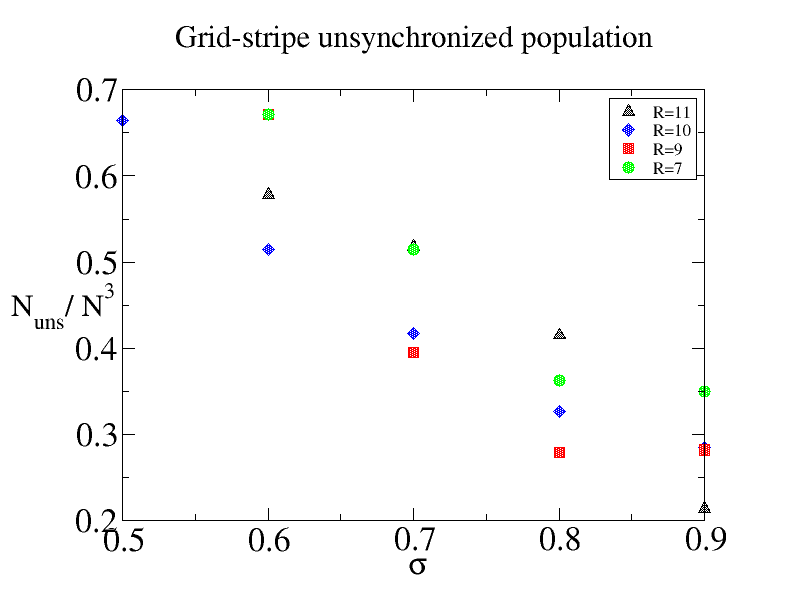}
\caption{\label{grid-inc-pop} (Color online) Relative number of unsynchronized neurons for cylinder-grid chimera states. Other parameters as in Fig.~\ref{fig1}.}
\end{figure}

The second observation concerns the cylinder-grid chimera state. By increasing \(\sigma\) 
we found that the number of unsynchronized neurons decreases. The number of
unsynchronized cylinders that run through the network stays the same but their 
diameter decreases. This is true for \(R=7,9,10,11\) as seen in Fig. \ref{grid-inc-pop}.
Both observations are in line with the results of the 2D system investigated in \cite{Theory_2D_Kasimatis2017}.

\section{General Conclusions and Future Work}
\label{sec:General Conclusions Future Work}

In this work we used a 3D LIF model to simulate the dynamics of neurons in a three dimensional non locally coupled network.
We focused on the identification and categorization of the stable chimera states that were observed. It was found that many chimera patterns are 
direct generalizations of chimeras observed in the 2D network \cite{Theory_2D_Kasimatis2017}: 2D spot \(\rightarrow\) 3D sphere, 2D single and multiple
layers \(\rightarrow\) 3D single and multiple layers, 2D grid \(\rightarrow\) 3D cylinder-grid, while others such as the cylindrical chimeras and cross- layered chimeras do not have a counterpart on the 2D system. Another observation was that chimera characteristics
such as the number of synchronized and unsynchronzied neurons,
the mean phase velocity range and the distribution profile differed significantly for coupling constant values greater than \(0.5\) and lower than \(0.4\) 
thus leading to the hypothesis that a critical point of the system exists between these values. For greater coupling constants
multistability was much more frequent than in the lower coupling constant side. Also, for lower coupling constants the unsynchronized areas of the 
chimera patterns exhibited higher mean phase velocities than the synchronized areas, while the opposite was true for the greater coupling constant.

\quad The aim of this work was to extend the 1D and 2D dimensional problems
to three dimensional networks with complex connectivity 
inspired by specific parts of the brain. It would be interesting to study the implementation of more complex 3D connectivity schemes in accordance with
natural connectivity patterns in the brain. To approach closer to the natural networks in the brain the replacement of periodic with specific boundary 
conditions is necessary and the calibration of the parameters to approach the natural neurons as close as possible needs to be addressed. All these 
efforts point toward the direction of understanding the flow of information in healthy brains, the understanding of brain functions and ultimately the 
confrontation with brain malfunctions and with neurodegenerative disorders.

\section{Acknowledgments} This work was supported by  computational
time granted from the Greek Research \& Technology Network
(GRNET) in the National HPC facility ARIS under  Grant
No. PR003017. T. Kasimatis acknowledges
support by Deutsche Forschungsgemeinschaft (DFG) in the
framework of the Collaborative Research Center 910 for
traveling expenses.
J. Hizanidis acknowledges support by the Ministry of
Education and Science of the Russian Federation in the
framework of the Increase Competitiveness Program of NUST
``MISiS'' (Grant No. К3-2017-057).

% \clearpage

\appendix
\section{Mean phase velocity gaps for all values of the coupling range.}
\label{sec:Appendix}

In Sec. \ref{sec:Quantitative analysis}, Fig. \ref{dw_wmaxmin_sample}, the maximum difference \(\Delta\omega_{max}\) in the mean phase  
velocities versus $\sigma$ were depicted for three different values of \(R\). In this Appendix we show the complete table for all the values of \(R\),
in order to demonstrate the change in the behavior of \(\omega\)-values around \(\sigma=0.4-0.5\).
The gap in the $\Delta \omega_{max}$ plots is associated with a transition taking place for $\sigma \ge 0.5$, see also Fig. 2. 

\begin{figure}[ht!]
\includegraphics[clip,width=1\linewidth,angle=0]{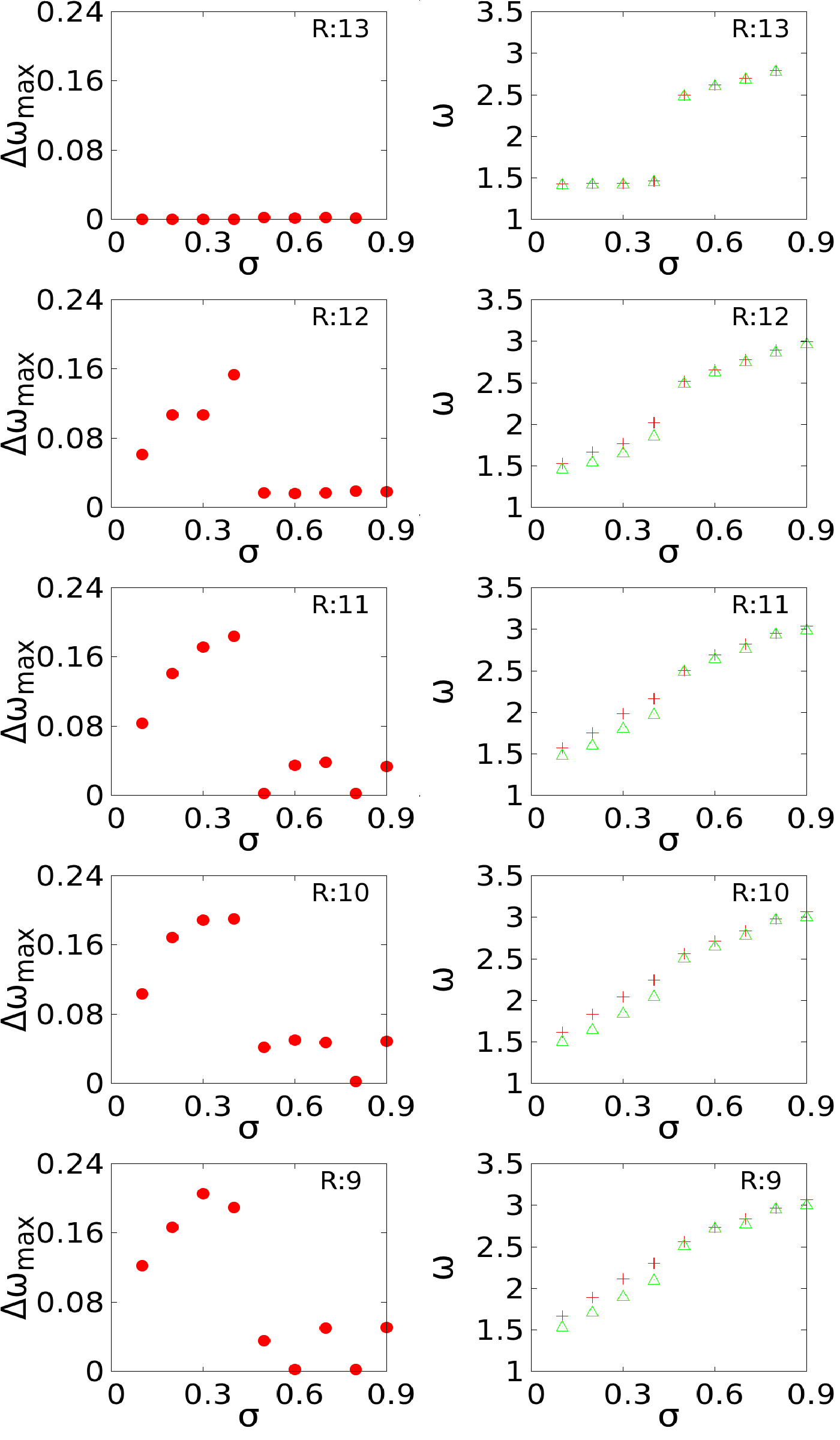}
\caption{\label{dwA} (Color online). The complete tables of maximum-minimum mean phase velocities and mean phase velocity range vs \(\sigma\) are depicted for all \(R\) values \(13-1\) . }
\end{figure}
\begin{figure}[ht!]
\includegraphics[clip,width=1\linewidth,angle=0]{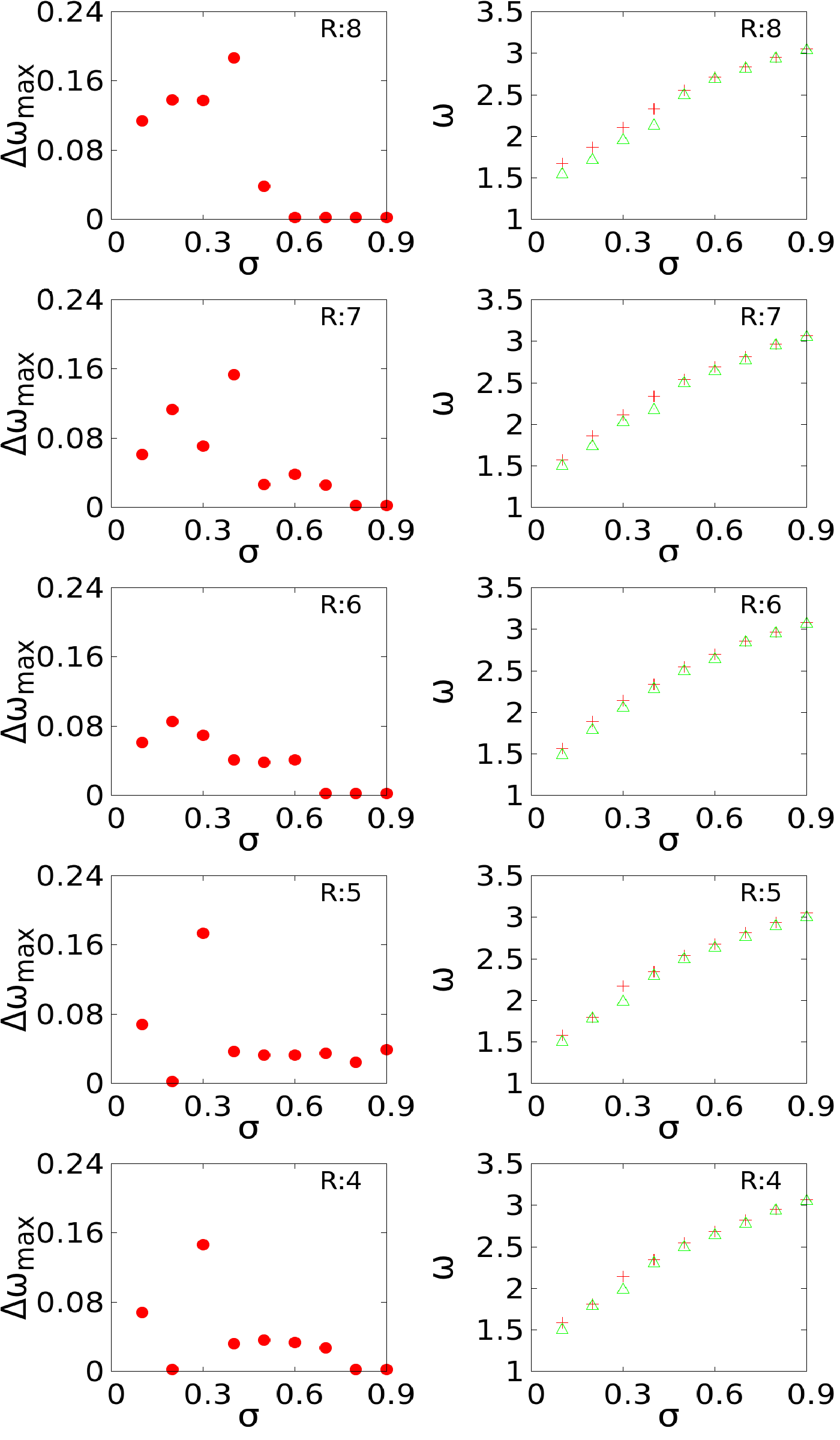}
%\caption{\label{dwB} (Color online). }
\end{figure}
\begin{figure}[ht!]
\includegraphics[clip,width=1\linewidth,angle=0]{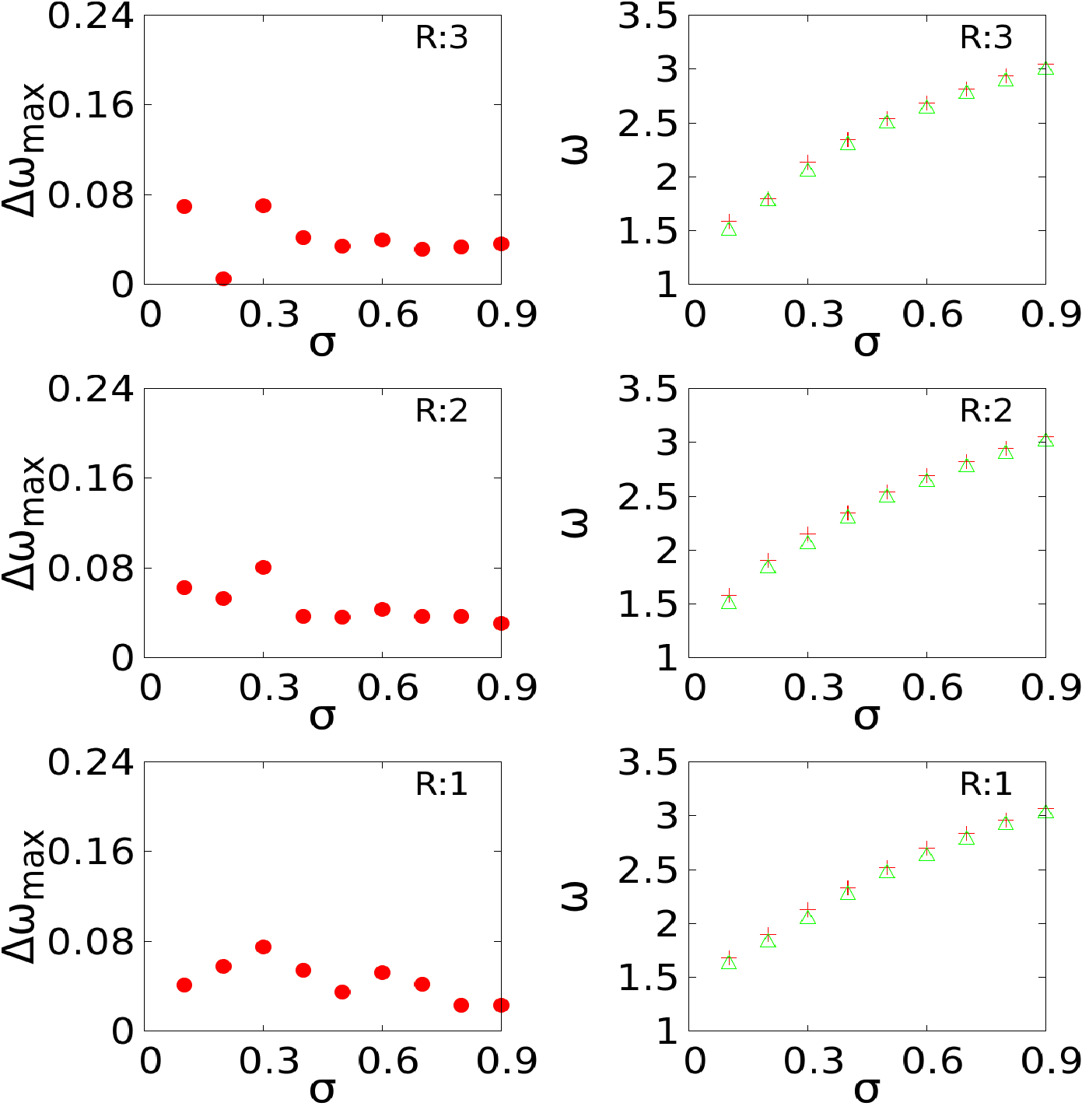}
%\caption{\label{dwC} (Color online). }
\end{figure}

 \clearpage

% \begin{thebibliography}{10}

\end{document}